\documentstyle[prb,aps,epsfig]{revtex}
\begin{document}
\title{Symmetry and designability for lattice protein models}
\author{Tairan Wang$^{1,2}$, Jonathan Miller$^1$, 
Ned S. Wingreen$^1$, Chao Tang$^1$\thanks{Corresponding author;
electronic mail: tang@research.nj.nec.com}, and Ken A. Dill$^3$}
\address{$^1$ NEC Research Institute, 4 Independence Way, Princeton, 
New Jersey 08540\\
$^2$ Department of Physics, Massachusetts Institute of Technology,
Cambridge, Massachusetts 02139\\
$^3$ Department of Pharmaceutical Chemistry, University of California,
San Francisco, California 94143}

\maketitle

\begin{abstract}
Native protein folds often have a high degree of symmetry.  
We study the relationship between the symmetries of native proteins, 
and their {\it designabilities} -- how many different sequences 
encode a given native structure.  Using a two-dimensional 
lattice protein model based on hydrophobicity, we find that those native 
structures that are encoded by the largest number of different sequences 
have high symmetry.  However only certain symmetries are enhanced, 
{\it e.g.} x/y-mirror symmetry and $180^o$ rotation, while others are 
suppressed.  
If it takes a large number of mutations to destabilize the native
state of a protein, then, by definition, the state is highly
designable. Hence, our findings imply that insensitivity to 
mutation implies high symmetry.
It appears that the relationship between 
designability and symmetry results because protein substructures are 
also designable. Native protein folds may therefore be symmetric
because they are composed of repeated designable substructures.
\end{abstract}


\section{Introduction}
The folded structures of proteins are often highly ordered. They are 
comprised of secondary structures, and often have striking regularities 
in their tertiary organization \cite{Levitt76,Richardson76}. 
What is the origin of symmetry in natural proteins? 

We approach this question by exploring the symmetries in simple lattice 
models of protein folding. Lattice models for proteins have been a rich
source of information on protein structure. Yue and Dill \cite{Yue95}
observed certain protein-like secondary structures and tertiary
symmetries in HP lattice model proteins that have low degeneracies, i.e.,
a small number of low energy states. More recently, Li et al. \cite{Li96}
noticed that the most ``designable" structures, namely those with a
large number of sequences folding into them, also often have global
symmetries. Since the most designable structures also have other
protein-like properties -- they have sharp thermal folding transitions and
are fast folders \cite{Melin99}, the connection to symmetry is
intriguing.

In these earlier studies, no quantitative measure was used to define
symmetry. Here, we explore in detail the connection between designability
and global symmetry, based on a quantitative, but simple, measure of
symmetry. Within the hydrophobic model \cite{Li98}, we quantify the
relation between designability and symmetry for 6x6 compact lattice proteins.

This article is organized as follows: Sec. II reviews
the hydrophobic model and the designabilities of structures.
In Sec. III, we relate symmetry to designability and
identify the importance of the surface-core pattern to
the particular emerging symmetries. To understand the origin
of enhanced symmetry, in Sec. IV we explore, first,  
the role of surface-to-core transitions
and, second, the extent to which symmetric folds
result from the repeated use of designable substructures.
For comparison, Sec. V addresses symmetry in a model not based on 
hydrophobicity. Sec. VI is the summary and conclusion.

\section{Hydrophobic Model}

In this section, we review the 
hydrophobic model and the designabilities of structures.
For more details, the reader is referred to Li et al. \cite{Li98}.

The hydrophobic model is a combination of the HP model \cite{Dill85}
and the solvation model cite{Eisenberg86}. 
In an HP model, the twenty different amino acids of proteins 
are replaced by two monomer types, 
Hydrophobic or Polar, according to their affinities for
water. Each protein is therefore a sequence of H's and P's.
In a {\it lattice} HP model, the amino acids are restricted to 
fall only on the sites of a regular lattice, typically a square lattice
in two dimensions or a cubic lattice in three dimensions.
The allowed conformations are self-avoiding, and 
hence cannot visit a single lattice site more than once.

\begin{figure}
\centering
\epsfig{file=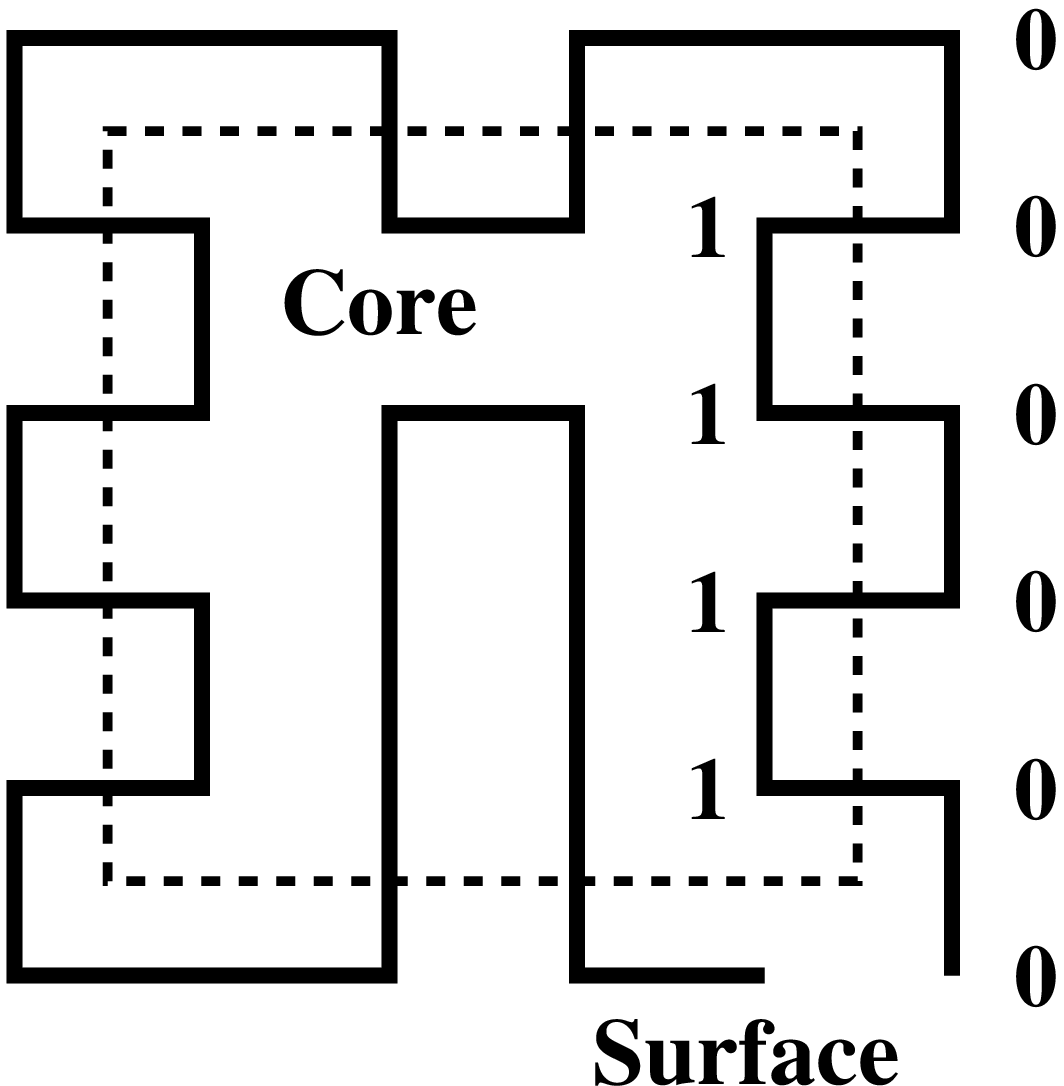,width=0.4\linewidth}
\vspace{1cm}
\caption{The most designable structure in the 6x6 hydrophobic
model. The 16 sites in the core region, enclosed by the dashed lines,
are represented by 1's; the 20 sites on the surface are represented by
0's. Hence the structure is represented by the string
001100110000110000110011000011111100.  The structure has an
approximate mirror symmetry, with an x-mirror symmetry score of 34,
i.e. with 34 bonds superposing upon reflection. The structure is also
highly ``pleated" with 12 surface-to-core transitions.}
\label{fig:fig.1}
\end{figure}

Here we use a variant that we call the hydrophobic model, in which 
only the maximally compact structures are considered as possible ground 
states.  This simplification still allows us to capture the 
essence of the HP model, but with two advantages: a substantial reduction 
in computational cost, and a conceptually useful method 
to represent sequences and structures within the same kind 
of abstract spatial representation, described below. In the hydrophobic 
model, the energy of a compactly folded protein is taken to be simply 
minus the number of H monomers in the ``core" (cf. Fig.~\ref{fig:fig.1}).
Therefore, in the hydrophobic model,
the energy of an HP-protein sequence folded into a particular compact
structure depends only on the structure's ordering of surface and core sites.  
Thus, a structure can be represented by a string ${\bf s}$ of 0's 
and 1's: sites in the core region are represented by 1's and sites
on the surface are represented by 0's, as illustrated in
Fig.~\ref{fig:fig.1}. 
Sequences are also represented by strings of 0's (P) and 1's
(H), ${\mathbf{h}} = (h_1, h_2, \ldots, h_N)$, where $h_i$ denotes the
hydrophobicity of the monomer at position $i$ of the sequence. 
The energy of a sequence folded into a particular
structure is therefore given by
\[
     H = - \sum_{i=1}^N s_i h_i
\]
where $s_i$ is the structure string. An equivalent way of writing the
energy is,
\[
     H = \frac{1}{2}\sum_{i=1}^N (s_i - h_i)^2 
          - \frac{1}{2}\sum_{i=1}^N s_i^2 
          - \frac{1}{2}\sum_{i=1}^N h_i^2. 
\]
The number of core sites is the same for all structures of
the same size, thus
$\frac{1}{2}\sum s_i^2$ is a constant and can be dropped.  
Similarly, the last
term, $\frac{1}{2}\sum h_i^2$, is a constant for each sequence and
so does not influence which structure is the ground state
for that sequence. Therefore, the only relevant term is the first
term, which measures the Hamming distance between the structure string
and the sequence string in an $N$-dimensional Euclidean space. A
sequence with string $\mathbf{h}$ will have a particular 
structure with string $\mathbf{s}$ as its unique ground
state if and only if $\mathbf{h}$ is closer to $\mathbf{s}$ than to
any $\mathbf{s'}$ corresponding to another structure.

\begin{figure}
\centering
\epsfig{file=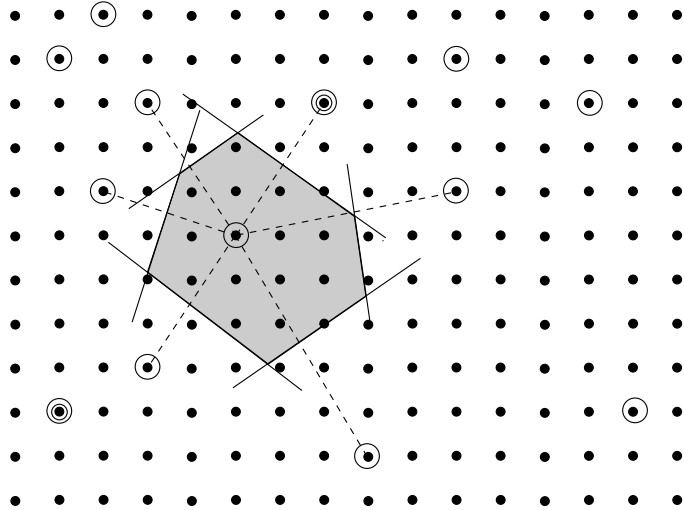,width=0.5\linewidth}
\vspace{1cm}
\caption{ Schematic representation of sequences and structures as
binary strings. Each vertex represents a possible sequence, {\it i.e.} a
string of 36 0's and 1's. Those vertices corresponding to structures
are circled. The sequences lying closer to a particular structure than
to any other have that structure as their unique ground state. The
designability of a structure is therefore the number of sequences
lying entirely within the Voronoi polygon about that structure. In
cases where more than one structure has the same string of 0's and
1's, {\it i.e.}, the same pattern of surface-core sites, the
corresponding
vertices are circled twice. These structures have zero
designability.}
\label{fig:fig.voronoi}
\end{figure}

The designability of a structure can therefore be obtained from the
following geometric construction: Draw bisector planes between
$\mathbf{s}$ and all of its neighboring structures in the
$N$-dimensional space. The volume enclosed by these planes is called
the Voronoi polytope around $\mathbf{s}$. The designability of a
structure is the number of sequences lying entirely within
the Voronoi polytope around that structure. This is schematically
represented in Fig.~\ref{fig:fig.voronoi}. Each vertex represents a
sequence. Those vertices corresponding to structures are
circled. Intuitively, the designability of a structure is closely
related to how far away its nearest neighbors are. The further 
away its neighbors are, the more designable it is.

\begin{figure}
\centering
\epsfig{file=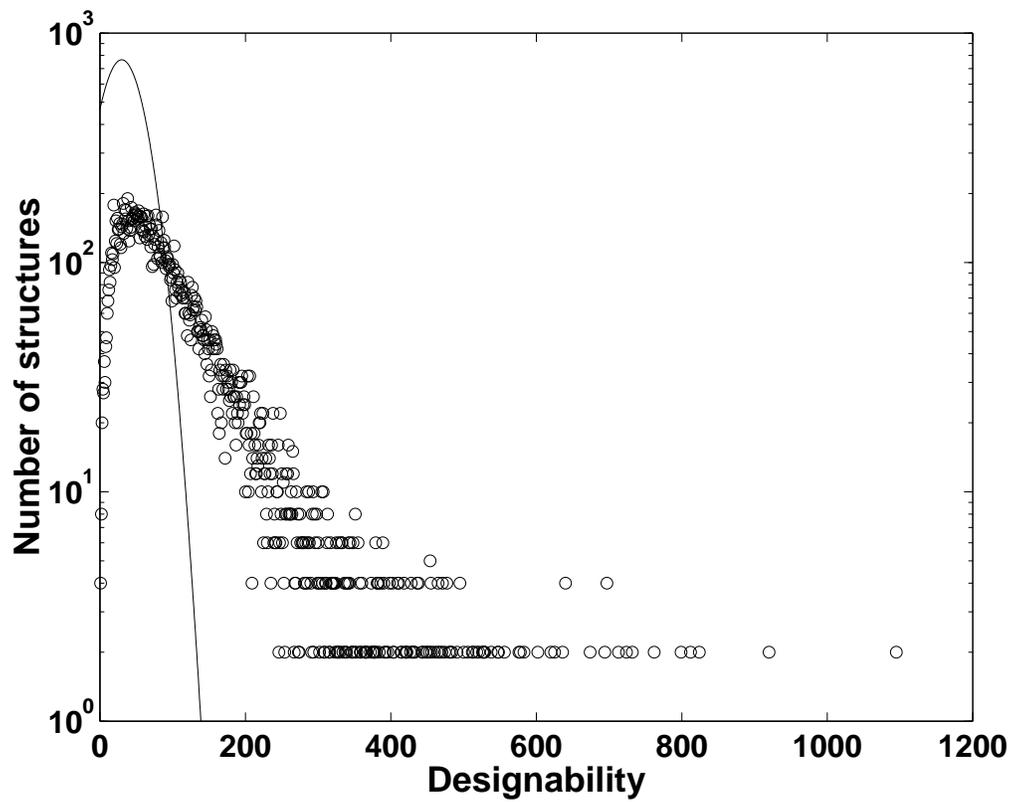,width=0.75\linewidth}
\vspace{1cm}
\caption{ Histogram of the number of structures versus designability
for the 6x6 hydrophobic model. The data is generated by sampling $20
\times
10^6$ sequence strings. For comparison, the solid line
shows the Poisson distribution
with the same average designability.}
\label{fig:fig.design_hist_hydro}
\end{figure}

The histogram of the number of structures versus designability for 
the 6x6 hydrophobic model is shown in Fig.~\ref{fig:fig.design_hist_hydro}.
The distribution has a long tail of highly designable structures
compared to a Poisson distribution
with the same mean. If sequences were randomly assigned to 
structures, the resulting distribution of designabilities would
be Poisson. It is clear from Fig.~\ref{fig:fig.design_hist_hydro}
that the structures in the tail have anomalously high designabilities.  
That is, they are unique ground states of many more than their share 
of sequences.

The many sequences that have a particular highly designable
native structure are related to each other by point 
mutations \cite{Li98}. For the model we consider, a point mutation
is simply the replacement of a hydrophobic monomer ``1" by
a polar monomer ``0", or vice versa.  Often, many monomers
can be independently mutated without destabilizing the
native state \cite{Li96}. Therefore, the folding
of these sequences is relatively insensitive to mutations.
One can think of highly designable structures as those which
remain most stable under sequence mutations.

\section{Symmetry and Designability}

In Li et al. \cite{Li96}, 
it was noted that highly designable structures tend to
be highly symmetric, with global mirror symmetries as well as regular
local motifs. In this study, we explore the connection 
between designability and symmetry in detail.
We focus on 6x6 2D square-lattice proteins.

To measure the symmetry of a structure, we look at how well that 
structure is preserved under rigid global transformations. Specifically, 
the transformations are the mirror reflections about the x/y axes, 
the mirror reflections about the two diagonal directions, and $90^o$ 
and $180^o$ rotations. The symmetry scores for a given structure are
the number of overlapping bonds between that structure and each of
its transformed versions.  The maximum
possible symmetry score for a 6x6 compact structure is $35$.

\subsection{Hydrophobic Model with Centered Core}

\begin{figure}
\begin{minipage}[b]{.46\linewidth}
\centering
\epsfig{file=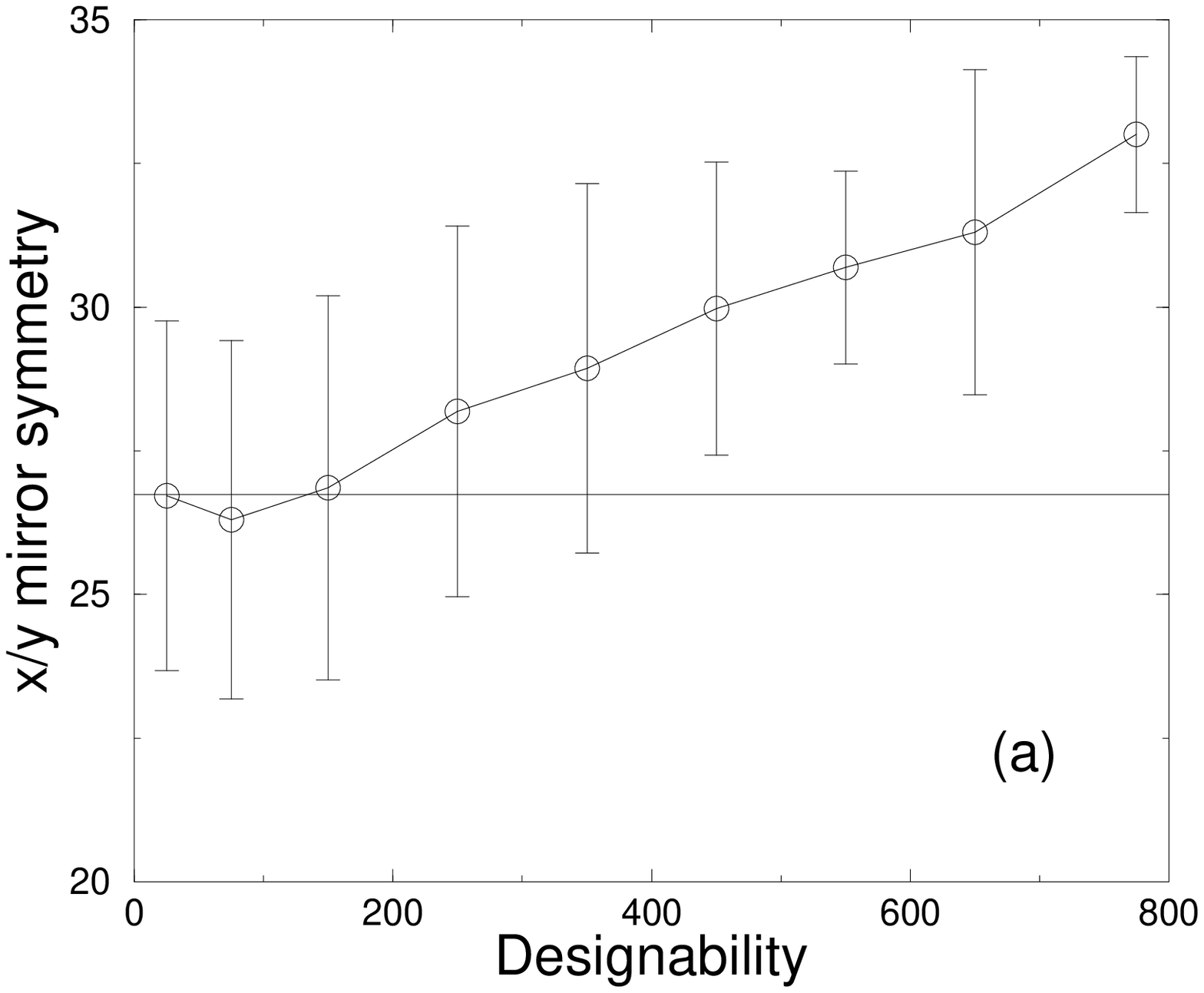,width=\linewidth}
\epsfig{file=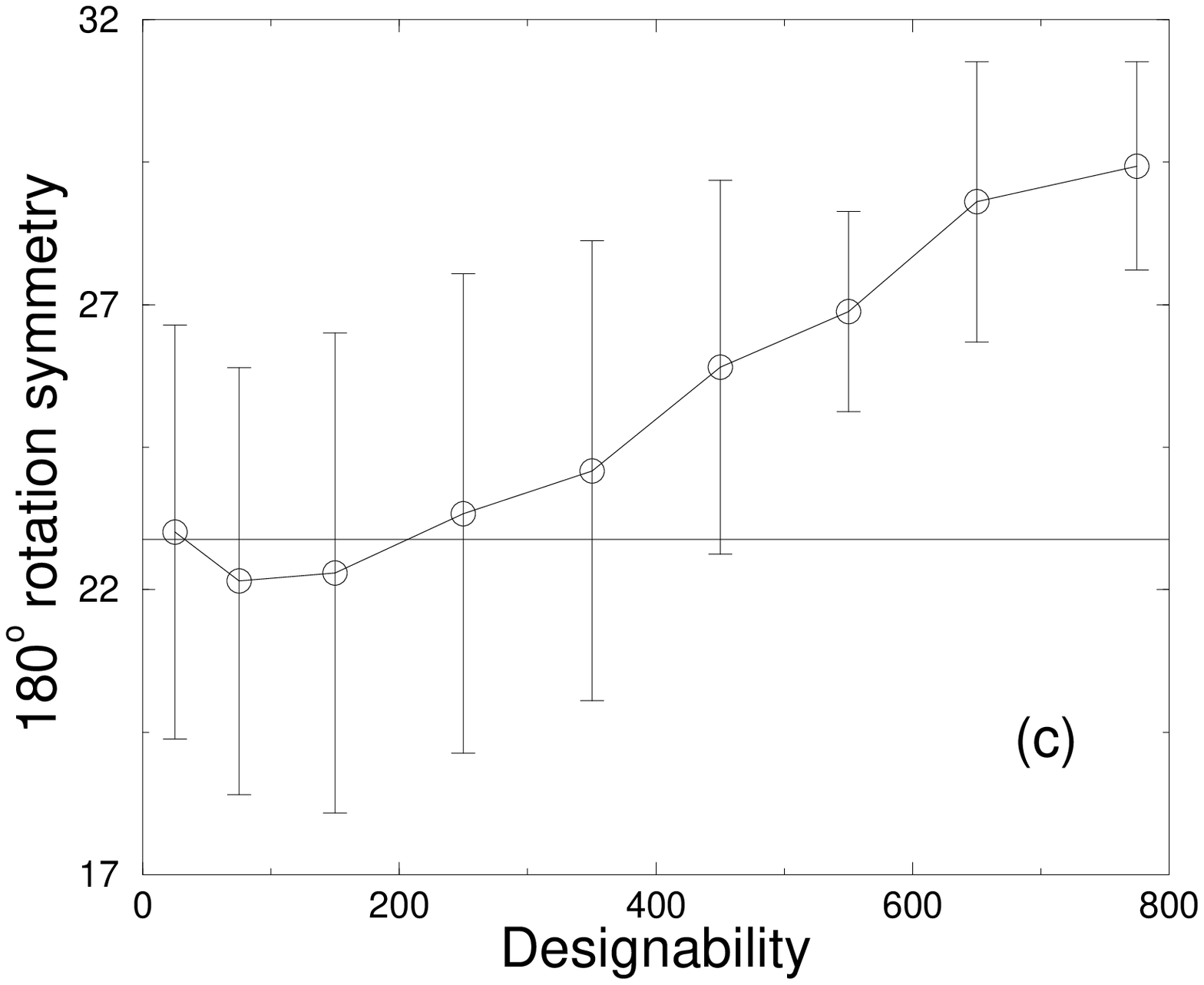,width=\linewidth}
\end{minipage}
\begin{minipage}[b]{.46\linewidth}
\centering
\epsfig{file=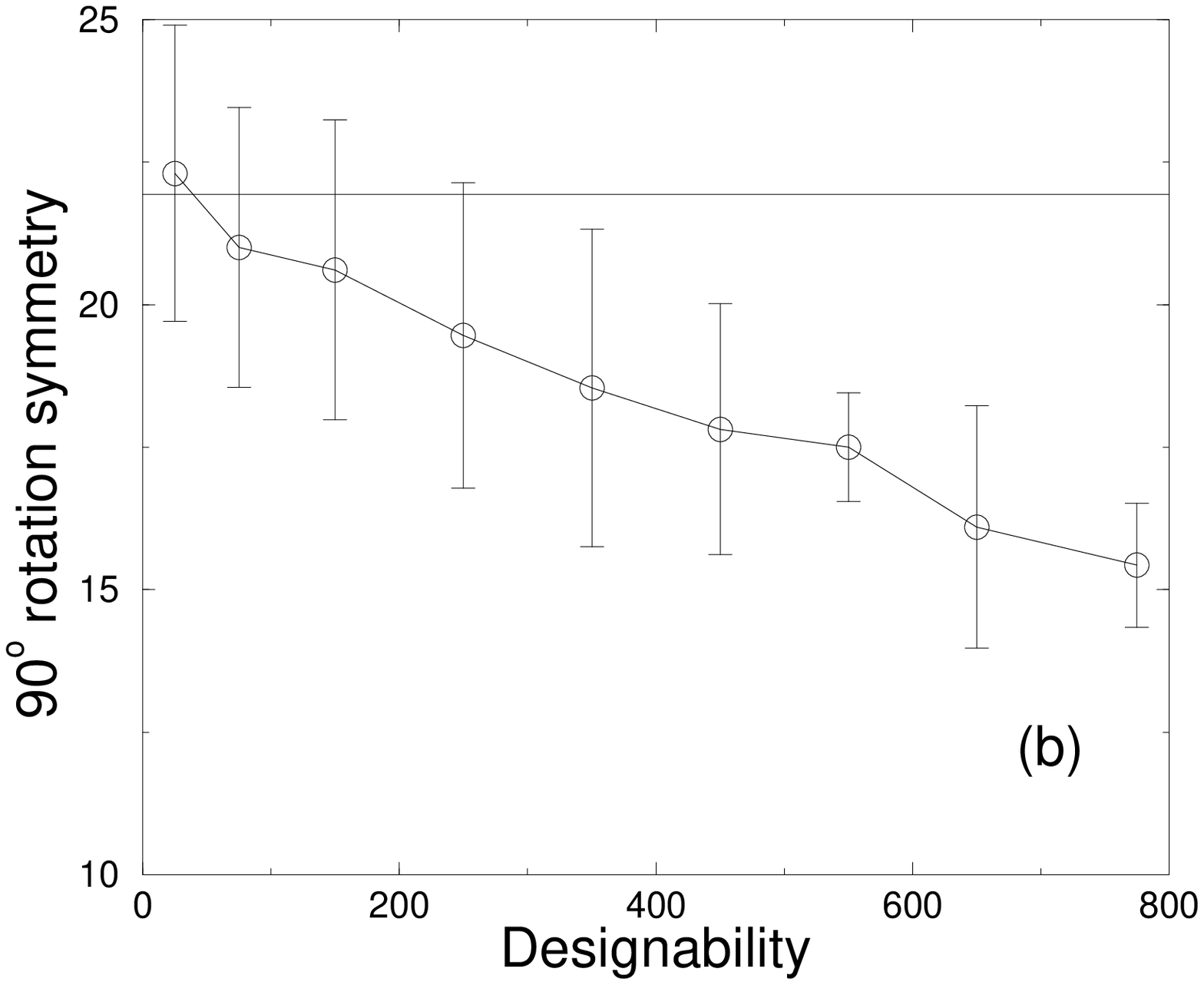,width=\linewidth}
\epsfig{file=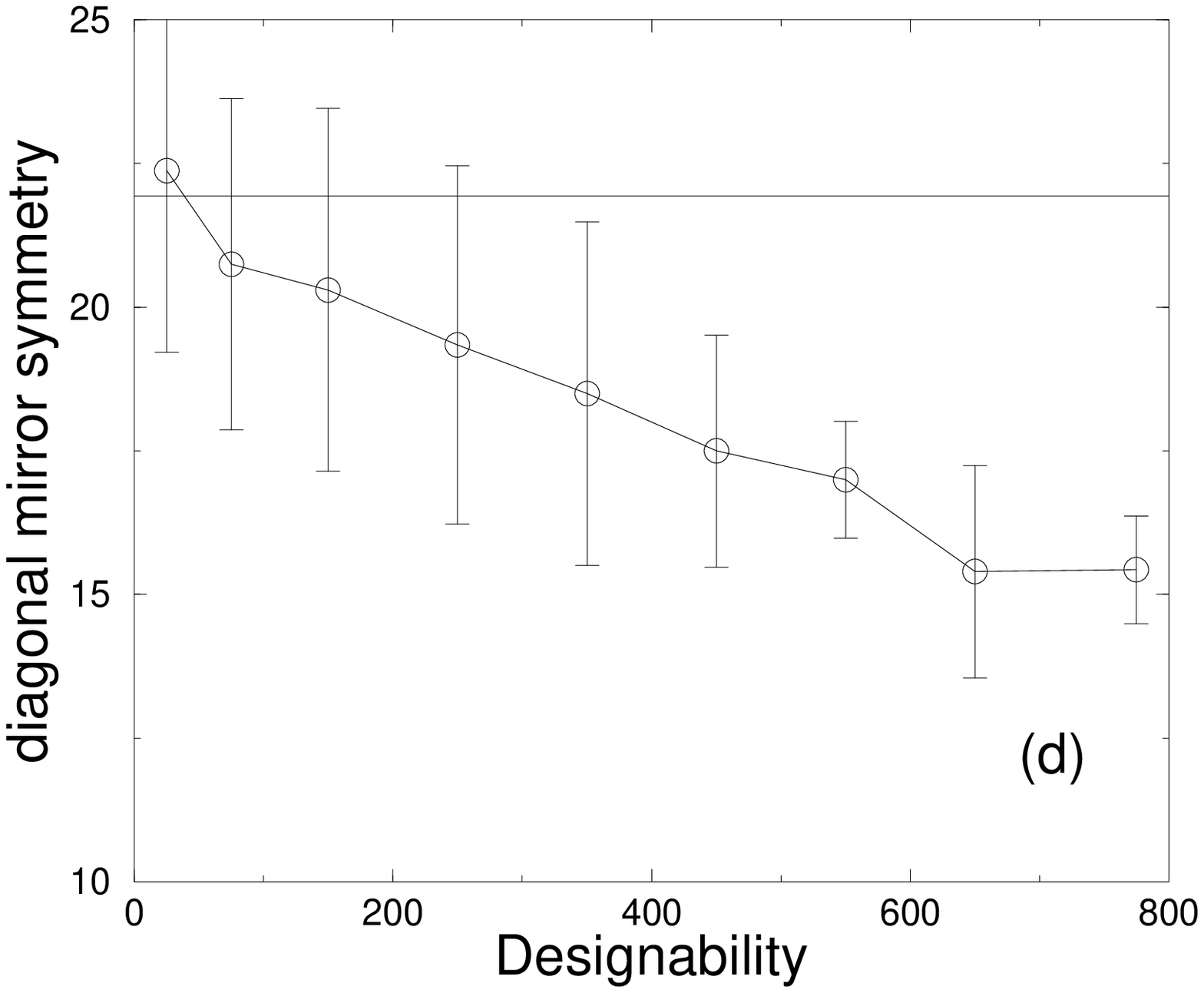,width=\linewidth}
\end{minipage}
\vspace{1cm}
\caption{ Averaged symmetry scores versus designability for the
hydrophobic model. The data is collected into bins according to
designability. The circles are the average symmetry scores within a
designability bin, and the error bars are the standard deviations. The
horizontal lines indicate the overall averaged symmetry scores in each
case. The x/y-mirror symmetry and the $180^o$ rotation symmetry
increase with designability, as shown in panels (a) and (c).  The
other symmetries decrease with designability, as shown in panels (b)
and (d).  }
\label{fig:fig.design_symm_hydro}
\end{figure}

We begin by studying the trends of symmetry versus
designability for the hydrophobic model. 
The symmetry scores,
averaged over designability bins, are plotted versus the designability
in Fig.~\ref{fig:fig.design_symm_hydro}.  It is observed that, on
average, the x/y-mirror symmetry 
(the larger of the x-mirror symmetry score and the
y-mirror symmetry score) increases with designability
(Fig.~\ref{fig:fig.design_symm_hydro}(a)). A similar trend is
observed for $180^o$ rotation symmetry, which is consistent with the
x/y-mirror symmetry result since a $180^o$ rotation is simply an x-mirror
operation followed by a y-mirror operation. For the other symmetry
operations, $90^o$ rotation and diagonal mirrors, the trend is
reversed -- higher designability implies lower symmetry scores for
these symmetries. Thus, for the hydrophobic model, there is indeed a
connection between designability and symmetry as previously noted
\cite{Li96}.
However, different symmetries behave differently with increasing
designability. In this case, the x/y-mirror symmetry and 
$180^o$ rotation symmetry are enhanced for
highly designable structures.

\subsection{Hydrophobic model with shifted core}

\begin{figure}
\centering
\epsfig{file=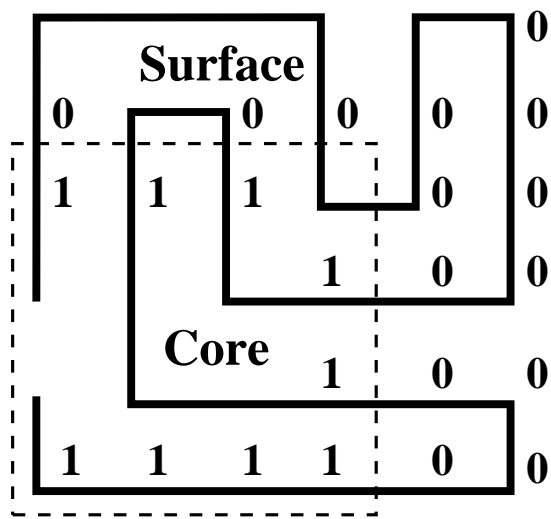,width=.4\linewidth}
\vspace{1cm}
\caption{Shifted-core hydrophobic model. The core region is shifted to
the lower left. Only one diagonal symmetry is still present in the
surface-core pattern.}
\label{fig:fig.hydro_shift}
\end{figure}

\begin{figure}
\centering
\epsfig{file=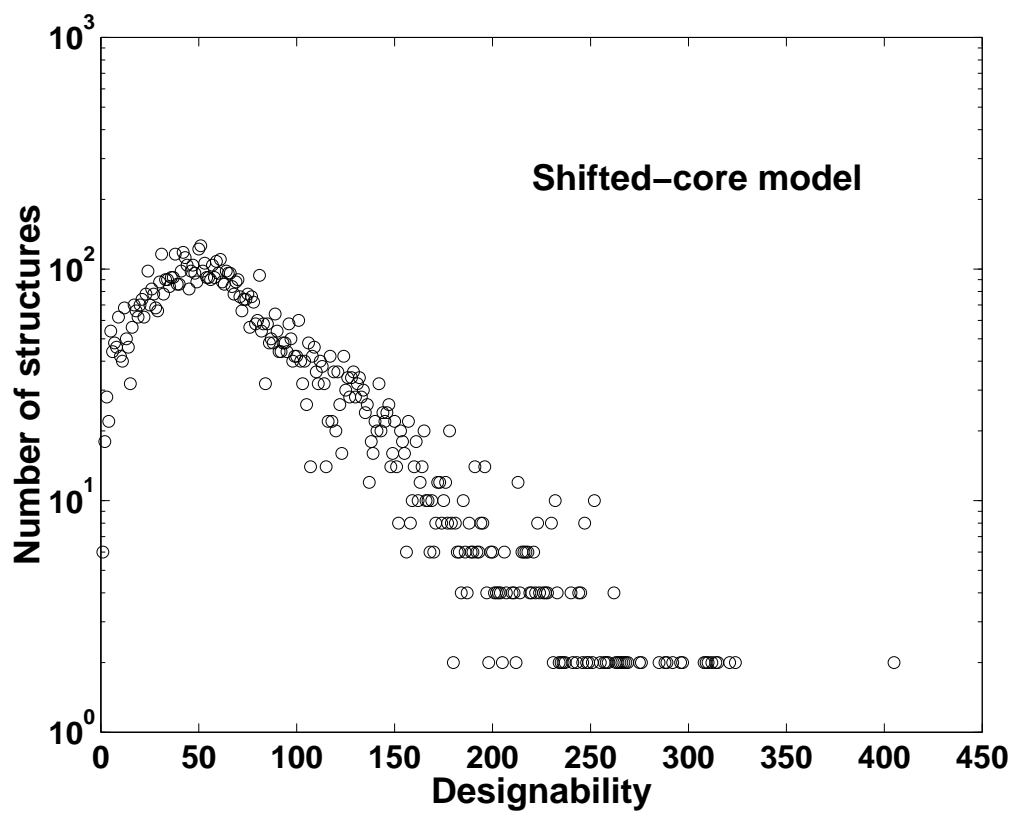,width=0.75\linewidth}
\vspace{1cm}
\caption{
Histogram of the number of structures versus designability for the
shifted-core model. The data is generated from a random sampling of
$6.3 \times 10^6$ sequence strings.
}
\label{fig:fig.design_hist_hydro_shift}
\end{figure}

For the hydrophobic model, the surface-core pattern has the symmetry 
of a square (Fig.~\ref{fig:fig.1}). What if this is not the
case? Does higher designability always lead to higher x/y-mirror
symmetry scores, even when the surface-core pattern is disrupted? To 
address this question, we study a shifted-core version of the hydrophobic
model. The core sites have been shifted to the lower corner
(Fig.~\ref{fig:fig.hydro_shift}).
In this shifted-core model, the energy of a compactly folded 
protein is taken to be simply minus the number of H monomers in 
the new off-center ``core" \cite{shiftedcore}.
The histogram of designability for the shifted-core model is shown in
Fig.~\ref{fig:fig.design_hist_hydro_shift}. Again, the region of high
designability is characterized by a long tail, qualitatively similar
to that of the ordinary 6x6 model. 

\begin{figure}
\begin{minipage}[t]{.45\linewidth}
\centering
\epsfig{file=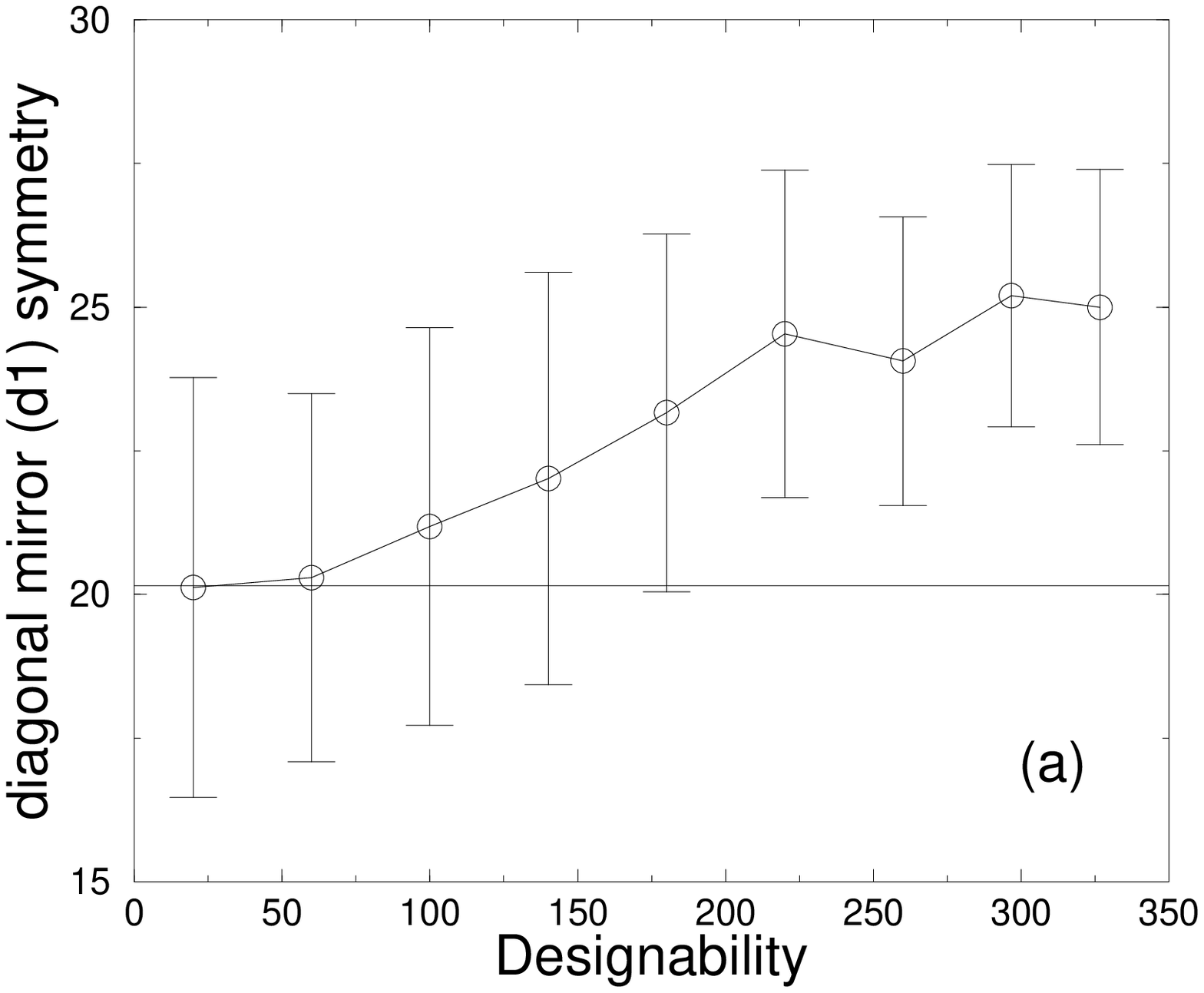,width=\linewidth}
\epsfig{file=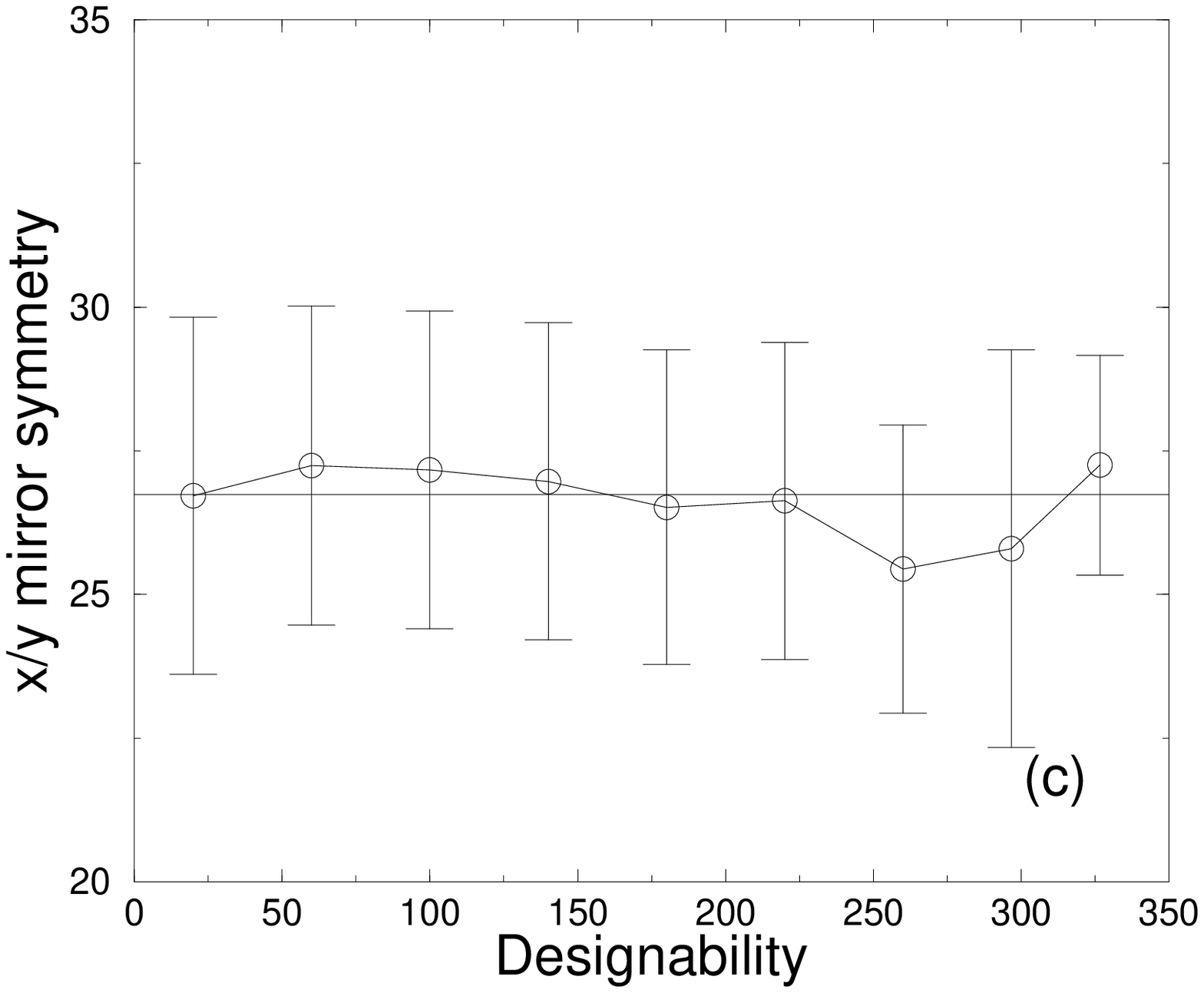,width=\linewidth}
\epsfig{file=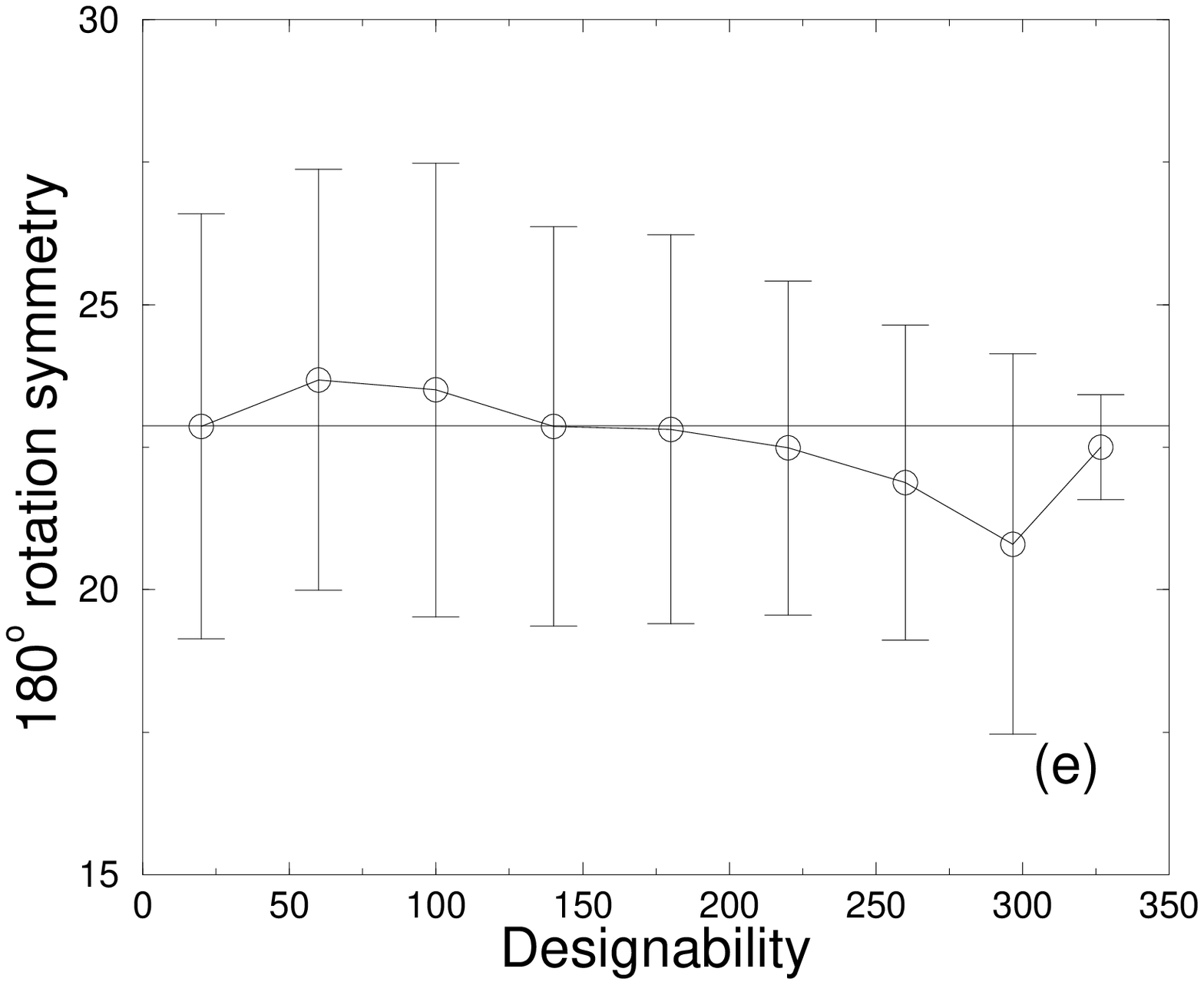,width=\linewidth}
\end{minipage}
\begin{minipage}[t]{.45\linewidth}
\centering
\epsfig{file=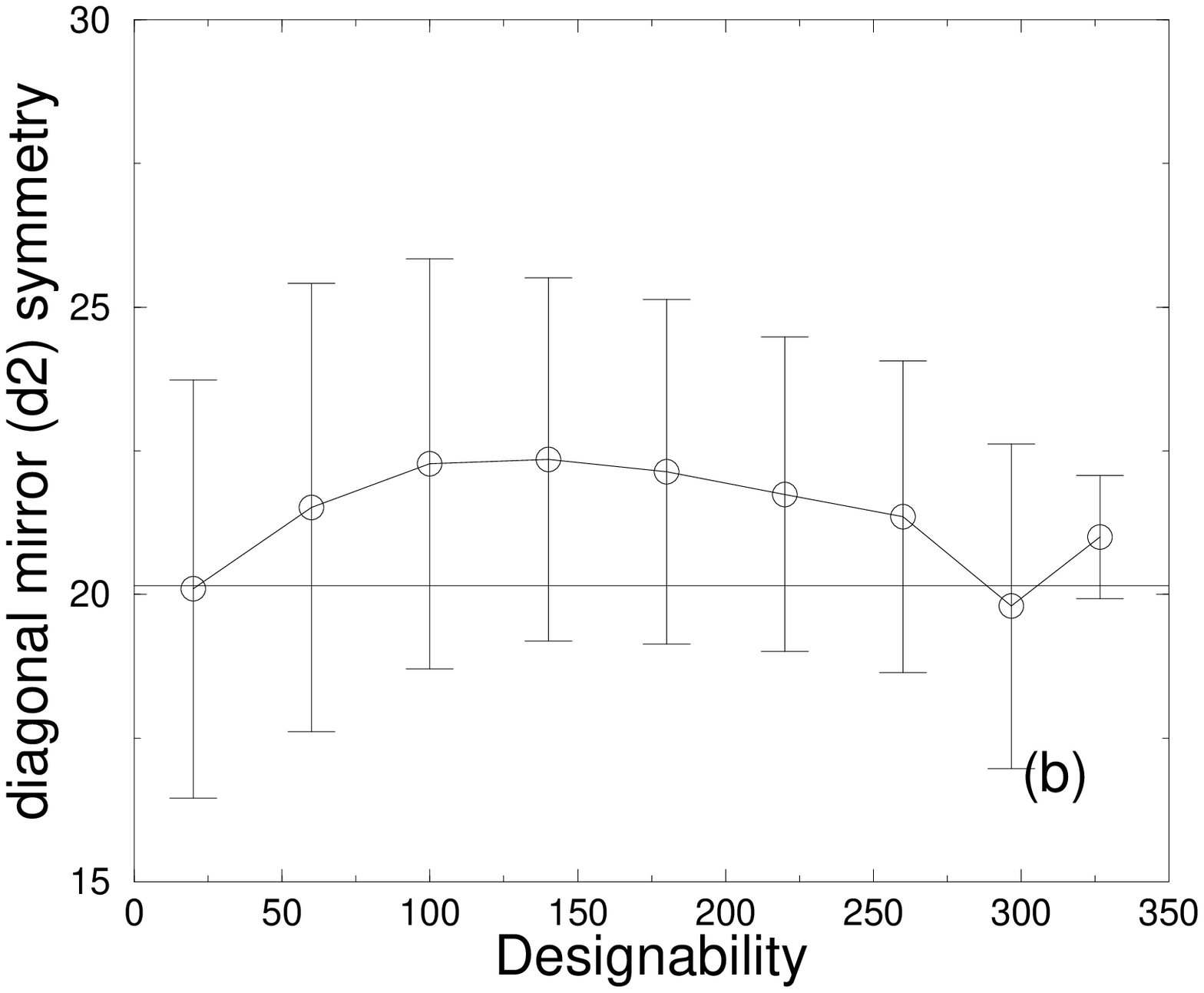,width=\linewidth}
\epsfig{file=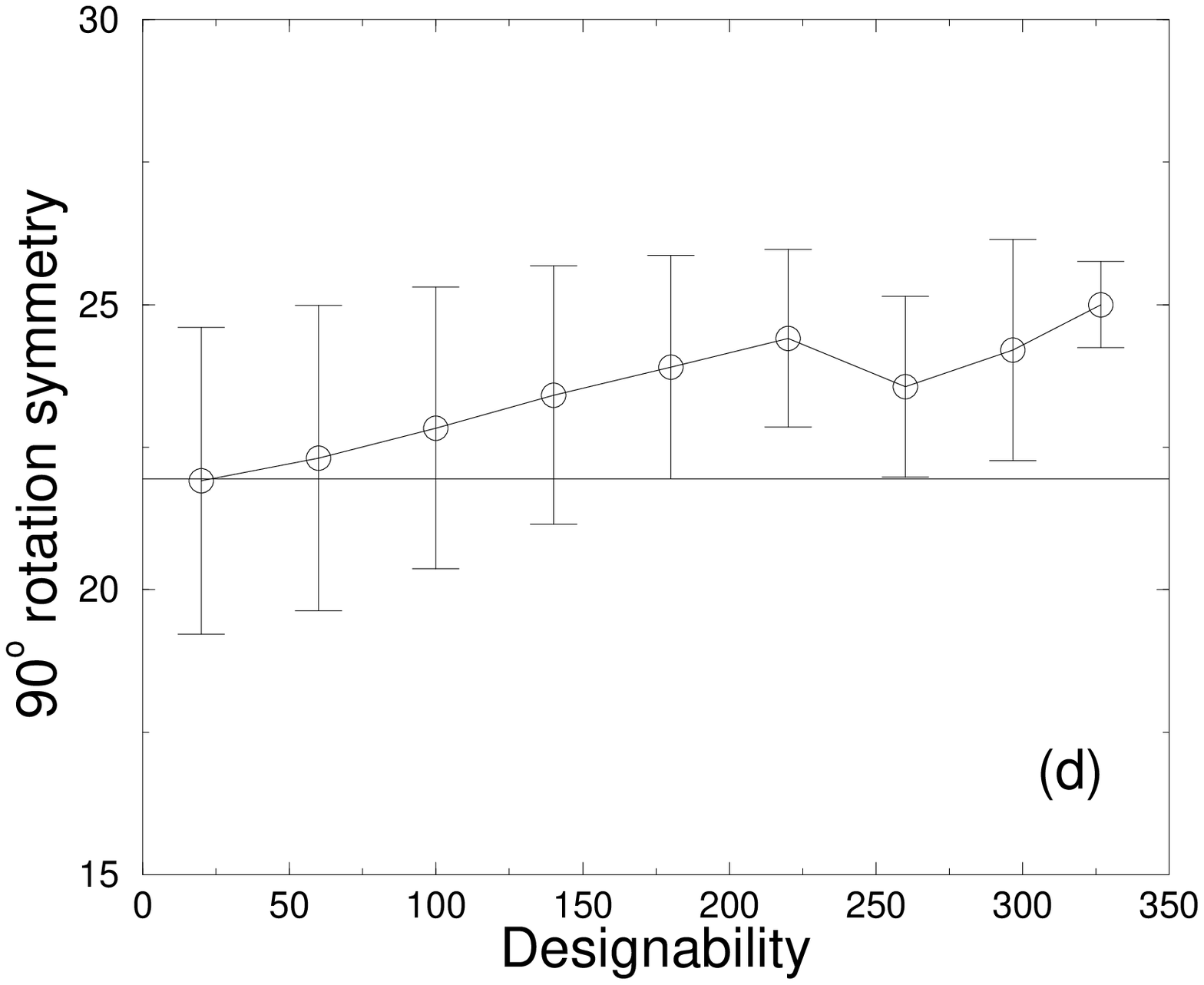,width=\linewidth}
\end{minipage}
\vspace{0.3cm}
\caption{
Averaged symmetry numbers versus designability for the shifted-core
model. ``d1'' is the diagonal-mirror symmetry preserved in the
surface-core pattern; ``d2'' is the other diagonal-mirror symmetry.
The horizontal lines indicate the overall average symmetry scores in
each case. The diagonal-mirror symmetry ``d1'' increases with
designability, while the other symmetries show little change.
}
\label{fig:fig.design_symm_hydro_shift}
\end{figure}

With the ``core" shifted to one corner, the surface-core pattern
no longer has mirror symmetries about the x and y axes.
In fact, only one diagonal mirror symmetry is left.
In Fig.~\ref{fig:fig.design_symm_hydro_shift}, we plot the averaged
symmetry scores versus designability for the shifted-core
model. When x/y-mirror symmetry is plotted versus designability
there is no correlation. Instead, only the diagonal-mirror
symmetry which is present in the surface-core pattern increases 
significantly with
designability (Fig.~\ref{fig:fig.design_symm_hydro_shift}(a)).

These results indicate that the surface-core pattern is
important in determining which symmetries are favored in
highly designable structures. In both cases considered,
the preferred symmetries follow the surface-core pattern.

\section{Origin of Symmetry}

Why is there an enhancement of symmetry for highly designable
structures? Also, why are some symmetries enhanced and not others? In 
this section, we
examine two possible origins of symmetry. First, perhaps global 
symmetries arise in designable structures because of a high number of 
surface-to-core transitions.  Or second, perhaps 
designable structures have global symmetries
because they arise from repeated highly designable substructures.

\subsection{Surface-to-core transitions}

A possible candidate for the link between designability and symmetry
is a local property of structures -- the number of surface-to-core
transitions. A surface-to-core transition occurs when monomer
$i$ of the chain is in the core and monomer $i+1$ is on the 
surface, or vice versa.
Previously \cite{Li98,Shih00},
it was observed that highly designable
structures have an excess of 
surface-to-core transitions. The connection can be
understood as follows: (1) A structure with a large number of 
surface-to-core transitions is difficult to rearrange without
exchanging many surface and core sites.
Such structures are therefore likely to be far from their
neighbors in the space of strings, and thus have 
a chance for high designability (cf. Fig.~\ref{fig:fig.voronoi}).
(2) In turn, a structure with a large number of surface-to-core transitions 
has a geometrical regularity which may naturally lead
to global symmetry. Moreover, the geometrical regularities,
and hence the enhanced global symmetry, should 
reflect the symmetry of the surface-core pattern, consistent
with our results using the shifted-core model. 

\begin{figure}
\centering
\begin{minipage}[b]{.4\linewidth}
\centering
\epsfig{file=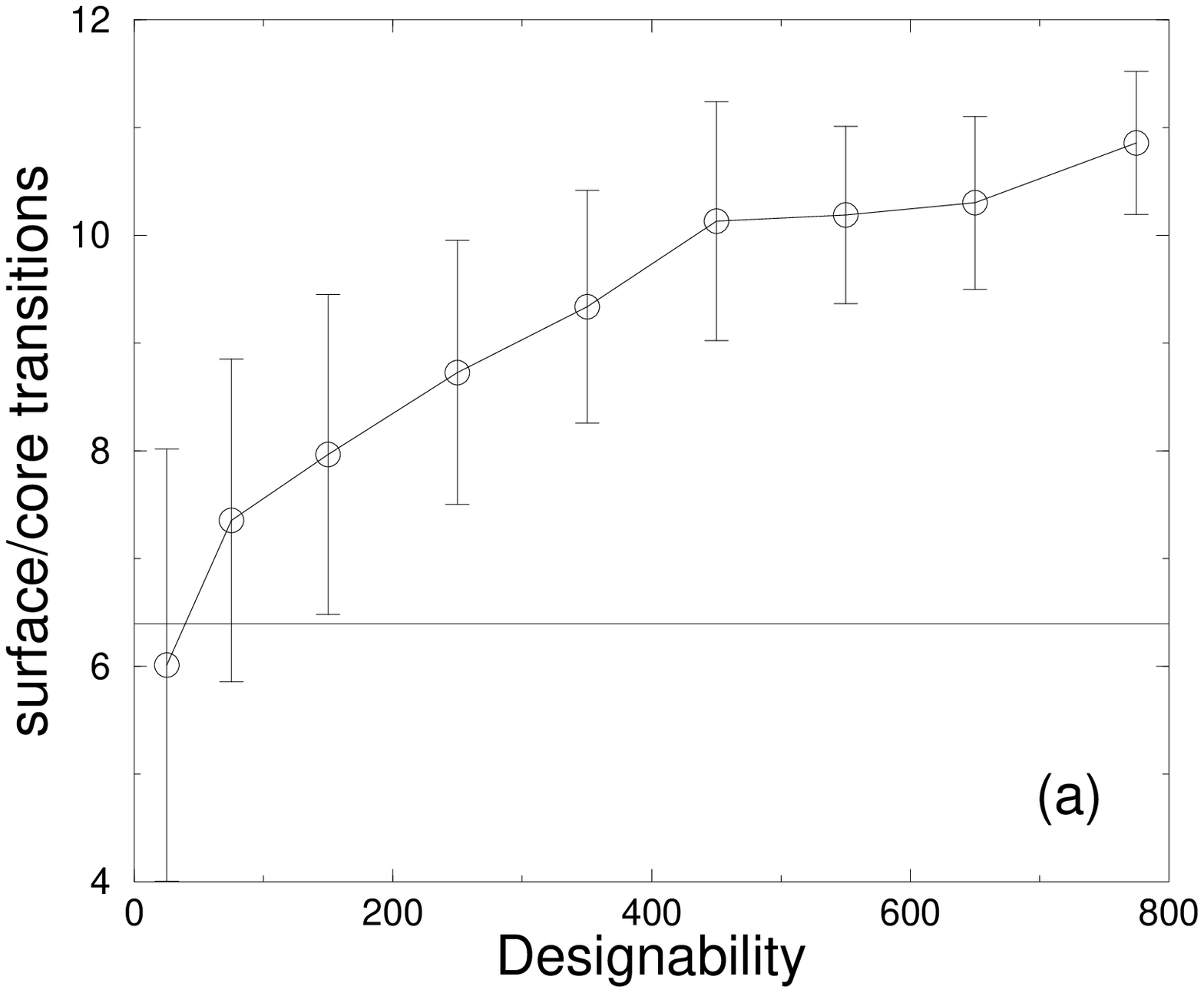,width=\linewidth}
\epsfig{file=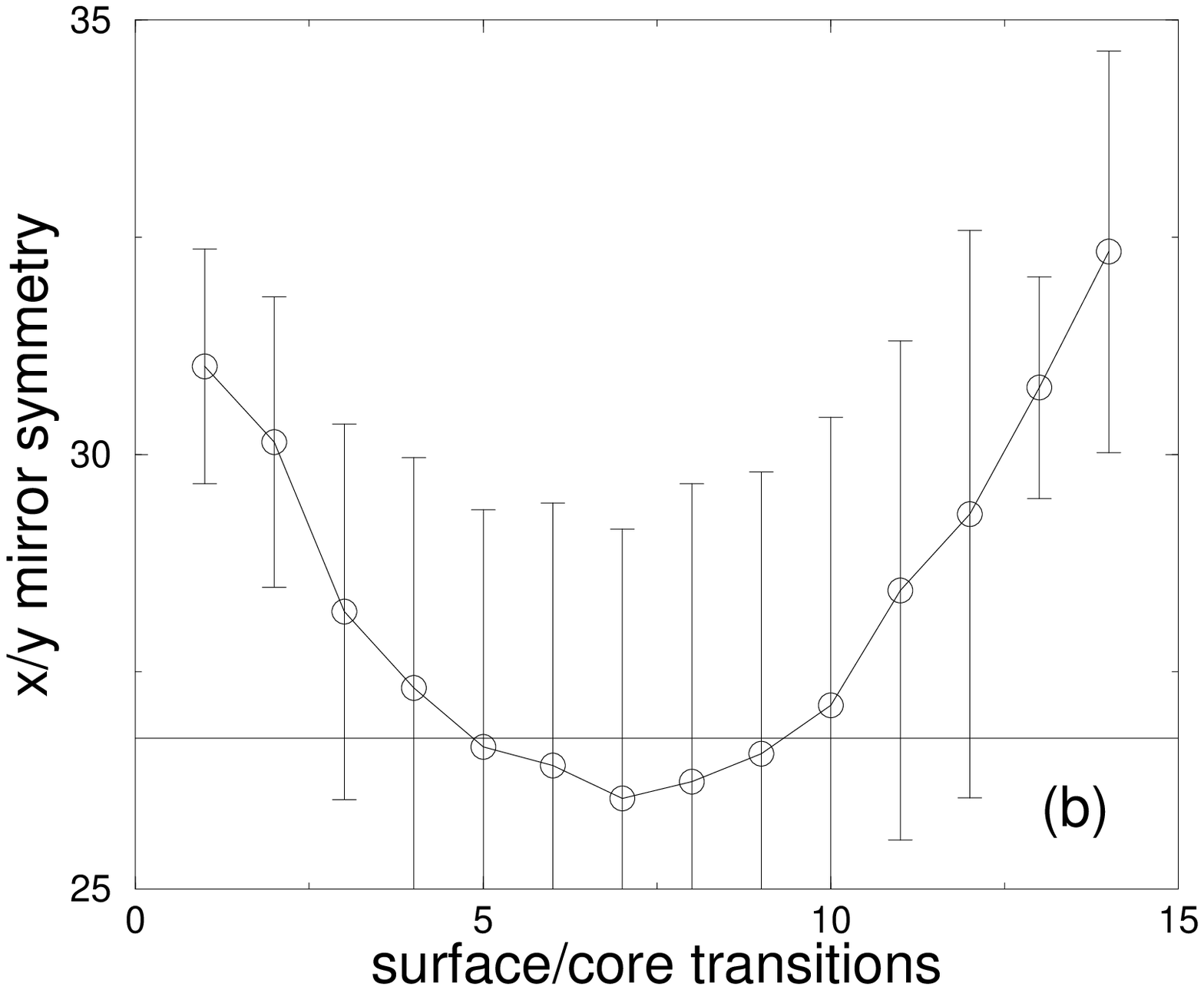,width=\linewidth}
\epsfig{file=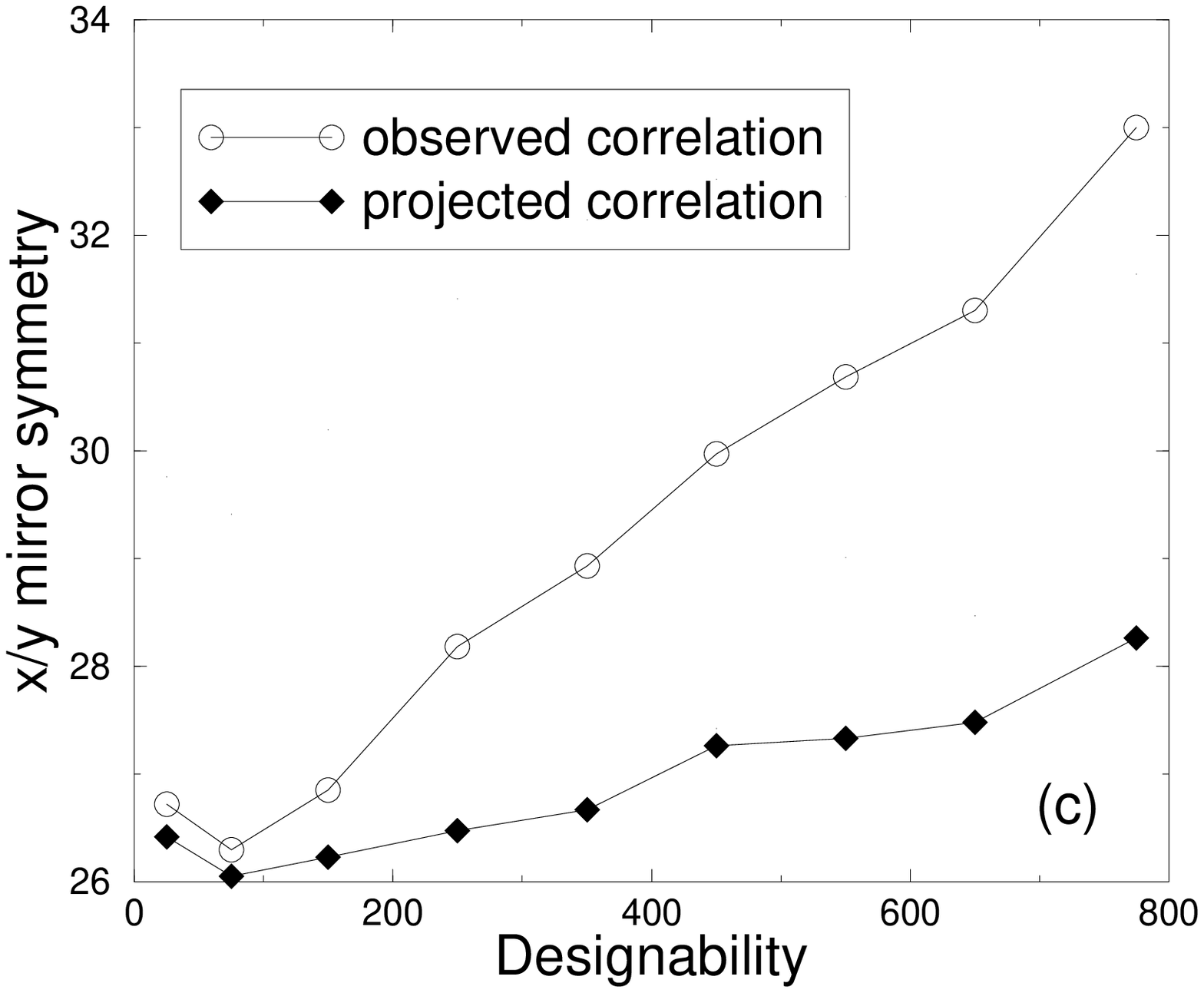,width=\linewidth}
\end{minipage}
\vspace{0.3cm}
\caption{
Panel (a) -- Averaged number of surface-to-core transitions versus
designability. The horizontal line indicates the overall averaged
number of surface-to-core transitions.
Panel (b) -- Averaged x/y-symmetry score versus number of surface-to-core
transitions. The horizontal line indicates the overall averaged
x/y-symmetry score.
Panel (c) -- Open circles indicate the averaged x/y-symmetry score
versus
designability; filled diamonds give the predicted x/y-symmetry score
versus
designability if we assume the connection between symmetry and
designability is only through the correlation of each with the number
of surface-to-core transitions.}
\label{fig:fig.projected}
\end{figure}

We tested whether surface-to-core transitions form
the link between designability
and global symmetry. We find that, qualitatively, both correlations
(1) and (2) are present, however, quantitatively, they
fail to account for the observed enhancement of global symmetry.
To quantify the first correlation, the number of surface-to-core transitions
averaged over structures of a given range of designabilities is
plotted against designability in Fig.~\ref{fig:fig.projected}(a)
for the original hydrophobic model.  High designability clearly
implies an enhanced number of surface-to-core transitions.
To demonstrate the second correlation, 
the x/y symmetry score averaged over structures with a given
number of surface-to-core transitions is plotted against the number of
surface-to-core transitions in 
Fig.~\ref{fig:fig.projected}(b). Symmetry does increase with the number of
surface-to-core transitions when the number of transitions is 
large \cite{lowstc}. 

Is the chain of correlations from designability 
to surface-to-core transitions to global symmetry strong
enough to explain the observed enhancement of global symmetry?
In Fig.~\ref{fig:fig.projected}(c), the x/y-symmetry score is plotted against
designability assuming the connection between them is only through the
correlation of each with the number of surface-to-core transitions.
Specifically, for a given designability, the corresponding average 
number of surface-to-core transitions is obtained from panel (a), then the
corresponding average x/y-symmetry score for that number of surface-to-core
transitions is obtained from panel (b).         
The predicted x/y-symmetry score thus
obtained is then plotted against designability. The actual x/y-symmetry
score versus designability from 
Fig.~\ref{fig:fig.design_symm_hydro}(a) is also plotted.  We see that
surface-to-core transitions account for only a fraction of the observed
connection between symmetry and designability.

\subsection{Designable substructures}

A second possible explanation for why designable folds are so symmetric 
is (1) they arise from designable substructures, and (2) 
symmetries are a natural consequence of assembling anything from 
identical substructures.

\begin{figure}
\centering
\epsfig{file=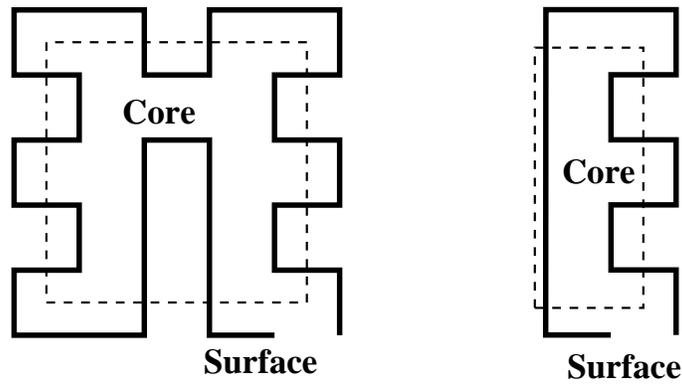,width=.5\linewidth}
\vspace{1cm}
\caption{The left half shows the most designable structure
for the 6x6 hydrophobic model; the right half shows the
most designable structure for a 3x6 hydrophobic model
which corresponds to one half of the 6x6 model.}
\label{fig:fig.sub}
\end{figure}

The most designable structure for the 6x6 hydrophobic model is shown
in the left half of Fig.~\ref{fig:fig.sub} . We take the right half of
the 6x6 surface-core pattern (a 3x6 rectangle) and calculate the
designabilities of all possible structures for this 3x6 hydrophobic
model. The most designable 3x6 structure is shown in the right half of
Fig.~\ref{fig:fig.sub}. Comparing the two structures, we see that the most
designable 3x6 structure is very similar to 
one half of the most designable 6x6 structure of the
original model. We conclude that this 6x6 structure is
highly designable because it is composed of two highly designable
substructures. 
The role of symmetry in this case can then be understood as
duplicating a winning solution. 

It is not yet clear how to quantify this concept of designable substructures.
Any scheme that involves breaking and reforming bonds,
as would be necessary to relate the structures in 
Fig.~\ref{fig:fig.sub}, seems arbitrary and unsatisfactory.
Nevertheless, a connection between designable
components and global symmetry seems to us likely,
and may have implications for understanding global
symmetries in real proteins.

\section{Symmetry beyond the hydrophobic model}

As a final question, we ask if the connection between designability
and symmetry is particular to models based on hydrophobicity, or
whether it occurs more generally. In the 
hydrophobic model each structure
is characterized by its ordering of surface and core sites.
As an alternative, we consider a model in which 
each structure is characterized by its complete contact matrix, as
described below.  Those structures with large contact-matrix 
distance from their neighbors are considered to be highly
designable. Within the contact-matrix model, the designable
structures {\it do not} show significantly enhanced symmetry.

Each structure has a contact matrix for its monomers. 
The elements of the contact matrix between monomers are 1 if they
are next to each other in the structure, but not adjacent on
the chain, and 0 otherwise.
A compact structure is uniquely defined by its contact matrix, 
up to rigid rotations and inversion. Thus, contact matrices and 
structures are related by a one-to-one mapping. 

\begin{figure}
\centering
\epsfig{file=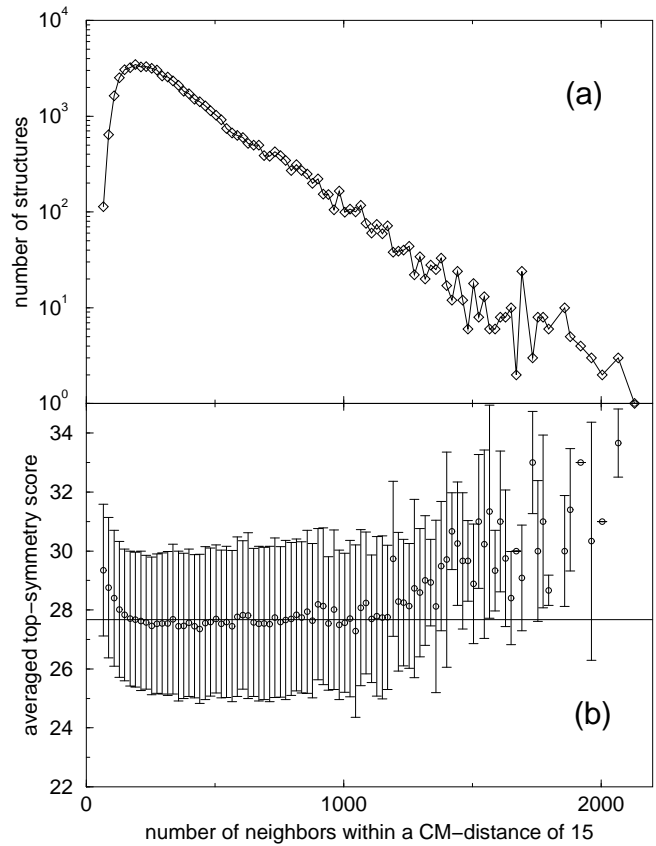,width=.5\linewidth}
\vspace{1cm}
\caption{
Panel (a) -- Histogram of the number of structures
versus number of neighboring structures within a Contact-Matrix
(CM) distance of 15. The maximum CM distance between any
two structures is 25. Panel (b) -- Averaged top-symmetry scores.
}
\label{fig:fig.symm_cmdist}
\end{figure}

The distance between any two structures can be measured by the overlap
between their contact matrices. The more overlap, the more bonds they
have in common. Hence, contact-matrix distance measures the similarity
of structures without particular emphasis on surface and
core sites. Just as in the hydrophobic model, where structures with
few neighboring strings emerged as highly designable, structures
with few neighbors in contact-matrix distance would emerge as
highly designable for more general models of amino-acid interaction
\cite{Goldstein00}.
For the set of compact 6x6 structures, we take the 
number of neighbors within a contact-matrix distance of
16 as a measure designability, with {\it low} number of neighbors 
implying {\it high} designability. The histogram of number of structures
versus number of near neighbors is shown in 
Fig.~\ref{fig:fig.symm_cmdist}(a).
Shown in Fig.~\ref{fig:fig.symm_cmdist}(b) is a plot of
the averaged top symmetry scores 
versus the number of neighboring structures within the
cutoff distance of $16$. The top symmetry score for a structure
is the highest score for all possible rotations and reflections.
The horizontal line indicates the average
top symmetry score of $27.7$. The region of few neighbors, and
hence high designability, is at the left of the figure
and has only a very slightly enhanced symmetry score with respect
to the average. 

We conclude that enhanced global symmetry of designable
structures {\it does not} emerge generally from models with 
arbitrary interactions among amino acids. Rather, the enhancement
of global symmetries is particular to models in which the
interaction between amino acids is dominated by hydrophobicity
\cite{Li96,Li98}. It appears that the correlation between the 
designability and symmetry of a native protein is a consequence of 
the key role played by hydrophobic solvation, and the approximate 
radial symmetries that result from it.

\section{Conclusion}

In this work, we have examined the connection
between the designability and symmetry of protein structures
within the hydrophobic model of Li et al. \cite{Li98}. 
The designable structures, namely those which are
unique ground states of many more than their share of sequences,
had been previously identified to have enhanced global symmetry,
as well as other protein-like attributes such as thermodynamic
stability and stability against mutations \cite{Li96}.
To quantify the relation between symmetry and designability
we focused on the set of two-dimensional compact 
structures which fill the sites of a 6x6 square lattice. 
We found that the designable structures have strongly
enhanced symmetry for x/y reflection and $180^o$ rotation.
For a related model in which the ``core" is shifted to one
corner, the only enhanced symmetry was a diagonal reflection.
This indicates that the enhanced symmetry of the designable
structures reflects a symmetry of the surface-core pattern.

To explore the origin of symmetry, we examined the
relation between designability, number of surface-to-core
transitions, and global symmetry. We conclude that an
increase in surface-to-core transitions among designable
structures can only account for a fraction of the observed
enhancement of global symmetry. Our working hypothesis is
that the global symmetry of designable structures results
from the repetition of designable substructures.

Finally, from a comparison model based on contact-matrix distances
we conclude that the relation between designability and symmetry 
originates from surface-core symmetries, which in turn, 
result mainly from hydrophobic interactions.

\newpage

\begin{thebibliography}{99}
\bibitem{shiftedcore}
The shifted-core model is not intended to be physically realistic.
It is, however, useful to probe the connection between symmetry and
designability.
\bibitem{lowstc}
Symmetry is also enhanced for structures
with minimal numbers of surface-to-core transitions, but these
represent only a small subset of the many structures with low
designability.
\end{thebibliography}
\end{document}